\documentclass[a4paper,pdflatex]{article}
\usepackage{graphicx,amsmath,amsfonts}

\def \ba{\begin{eqnarray}}\def\ea{\end{eqnarray}}
\def\bc{\begin{center}}\def\ec{\end{center}}

\title{\bf Moli\`ere's multiple scattering theory revisited}
\author{\bf O. Voskresenskaya\footnote{On leave of
absence from Siberian Physical Technical Institute. Electronic
address: voskr@jinr.ru} ~and A. Tarasov}
\date{}
\begin{document}

\maketitle
 \bc
Joint Institute for Nuclear Research, 141980 Dubna,
Russia \ec

\begin{abstract}
\noindent A part of Moli\`{e}re's multiple scattering theory
concerning the determination of the screening angular parameter is
revised. An universal form of the Coulomb corrections to the
screening angle, the exponential part of the distribution function,
and the angular distribution is discussed within the small-angle
approximation of this theory. The accuracy of the Moli\`{e}re theory
in determining the screening angle is estimated.

\end{abstract}

\section{Introduction}

The theory of multiple scattering of fast charged particles by atoms
is of importance in analysis of experimental results for many
high-energy experiments, such as \cite{Dirac} etc. Precise
measurements of multiple scattering effect in these experiments
requires  adequate accuracies in their theoretical description.

Multiple scattering of charged particles in the Coulomb field of
nuclei is described by a number of theoretical treatments [2--6].
The various theories differ mainly in their treatment of the single
scattering law. The Moli\`{e}re method is independent of the exact
form of the single scattering law, but contains a model-dependent
parameter representing the atomic screening, the so-called
`screening angular parameter' $\chi_a$, which enters into other
important quantities of the Moli\`{e}re theory.

Moli\`{e}re calculated his screening parameter by using the
Thomas--Fermi potential and the WKB method. He obtained an
approximate expression for this parameter
\ba\label{1} \chi_a\approx\chi_a^{\scriptscriptstyle
B}\sqrt{1+3.34\left(Z\alpha/\beta\right)^2}, \ea
valid to second order in $a=Z\alpha/\beta$, where only first term is
determined quite accurately, while the coefficient in the second
term is found numerically and approximately.

In the present work, we have obtained for $\chi^{}_a$ and some other
quantities of the Moli\`{e}re theory rigorous results valid in all
orders of the parameter $a$. In other words, we have found
analytical expressions for the so-called `Coulomb corrections' to
the Born results. Also, we have evaluated numerically these Coulomb
corrections and studied their $Z$-dependence. In addition, we have
estimated the accuracy of the Moli\`{e}re theory in determining the
screening angle $\chi_a$.

The outline of the paper is as follows. In Sections 2--4, we review
some basic results of \cite{M47}, i.e., solving the transport
equation (Sec. 2), Moli\`{e}re's expansion method (Sec. 3), and
determining the screening parameters by Moli\`{e}re (Sec. 4). The
results of the present work are given in Sections 5--6. In Sec. 5,
we consider an another determination of the screening parameters
allowing to obtain rigorous relations between their exact and Born
values. In Sec. 6, we evaluate the numerical values of the obtained
Coulomb corrections in the range $Z=4$ to $Z=82$. Also, we estimate
the accuracy of the Moli\`{e}re theory in determining the screening
angle. Finally, in Sec. 7, we summarize the main results of this
work. In Appendix, we present an alternative way of obtaining the
approximate solution of the transport equation for the thick
targets.

\section{The transport equation and its solution}

The basis for studies of multiple scattering effects
 in a nearly-isotropic and quasi-homogeneous medium by the transport
equation method is the Boltzmann transport equation often used in
statistical physics of systems with a large number of degrees of
freedom. It can be used as well in the relativistic Moli\`{e}re
scattering problem \cite{Bethe,Scott49} within the semiclassical
approach to particle transport in matter.

Let all scattering angles are small $\theta \ll 1$ so that $\sin
\theta\sim\theta$, and  $\sigma(\chi)$ be the elastic differential
cross section for the single scattering into the angular interval
$\vec\chi =\vec\theta-\vec\theta^\prime$. Define now
 $W_{\scriptscriptstyle M}(\theta,t)\theta d\theta$ as the number of scattered
 particles in the interval $d\theta$ after traversing a thin homogeneous
foil of thickness  $t$. Then can be used the standard transport
equation \cite{Bethe}:
\ba\label{kinet} \frac{\partial W_{\scriptscriptstyle
M}(\theta,t)}{\partial t}= -n_0 W_{\scriptscriptstyle
M}(\theta,t)\int \sigma(\chi)d^2\chi +n_0 \int W_{\scriptscriptstyle
M}(\vec\theta-\vec \chi,t)\sigma(\chi)d^2\chi,\ea
where $n_0=(N_A\rho)/M$ (cm$^{-3}$) is the number density with the
Avogadro number $N_A=6.02\times 10^{23}$ mol$^{-1}$, the mass
density of the target matter $\rho$ measured in units g/cm$^3$, and
the molar mass of target atoms $M$ (g/mole). The quantity $n_0$ is
the number of the target atoms per cm$^3$.

Following Moli\`{e}re, we introduce the  Fourier--Bessel
transformation of distribution and get to the distribution function
$W_{\scriptscriptstyle M}(\theta,t)$ a general expression
\ba\label{m1} W_{\scriptscriptstyle M}(\theta,t)=
\int\limits_0^{\infty} J_0(\theta \eta)g(\eta,t) \eta \,d\eta,\ea
in which
\ba \label{m2} g(\eta,t)=\exp[ N(\eta,t)-N_0(0,t)],\ea
$\theta$ is the polar angle between the track of a scattered
particle and the initial direction $z$, $\eta$ is the Fourier
transform variable corresponding to $\theta$, and the Bessel
function $J_0$ is an approximate form for the Legendre polynomial
appropriate to small scattering angles \cite{M47,Bethe}.

In the notation of Moli\`{e}re,
\ba \label{m3}N(\eta,t)= 2\pi n_0
t\int\limits_0^{\infty}\sigma(\chi)J_0(\chi\eta)\chi d\chi, \ea
and $N_0$ is the value of \eqref{m3} for $\eta=0 $, i.e., the total
number of collisions
\ba \label{m4}N_0(0,t)= 2\pi n_0
t\int\limits_0^{\infty}\sigma(\chi)\chi d\chi. \ea
The magnitude of $N_0-N$ is much smaller than $N_0$ for values
$\eta$, which are important. It can be called `the effective number
of collisions'.

Inserting \eqref{m2}--\eqref{m4} back into \eqref{m1}, we have
\ba\label{mol3} W_{\scriptscriptstyle M} (\theta,t)=
\int\limits_0^{\infty}\eta\, d\eta J_0(\theta \eta)\exp\left[-2\pi
n_0 t \int \limits_0^{\infty}\sigma(\chi)\chi d\chi[1-J_0(\chi
\eta)]\right]. \ea
This equation  is exact for any scattering law, provided only the
angles are small compared with a radian, and is equivalent to Lewis'
result \cite{GS}.

For $g(\eta,0)=1$ and all $\eta$, the expressions
\eqref{m1}--\eqref{m4} can be rewritten as follows:
\ba\label{mol1} W_{\scriptscriptstyle M}(\theta,t)=
\int\limits_0^{\infty} J_0(\theta \eta)e^{-n_{
 o}tQ(\eta)} \eta\, d\eta, \ea
where
\ba \label{mol2} Q(\eta)=2\pi \int\limits_0^{\infty}
\sigma(\chi)[1-J_0(\chi \eta)]\chi d\chi. \ea
This result is mathematically identical to the result of Snyder and
Scott for the distribution of projected angles \cite{Scott49}.

\section{Moli\`{e}re's expansion method}

One of the most important results of the Moli\`{e}re theory is that
the scattering is described by a single parameter, the so-called
`screening angle' ($\chi_a$ or $\chi_a^{\,\prime}$):
\ba\chi_a^{\,\prime}=\sqrt{1.167}\,\chi_a=
\left[\exp\left(C_{\scriptscriptstyle
E}-0.5\right)\right]\chi_a\approx1.080\,\chi_a, \ea
where $C_{\scriptscriptstyle E}=0.57721$ is the Euler constant.

More precisely, the angular distribution  $W_{\scriptscriptstyle
M}(\theta)\theta d\theta$ depends only on the logarithmic ratio of
the `characteristic angle' $\chi_c$ describing the foil thickness to
the `screening angle', which describes the scattering atom:
\ba b=\ln \left(\frac{\chi_c}{\chi_a^{\,\prime}} \right)^2\equiv\ln
\left(\frac{\chi_c}{\chi_a} \right)^2+1-2C_{\scriptscriptstyle
E}\sim \ln N_0\,. \ea

The screening angle $\chi_a$ can be determined approximately by the
relation
\ba\label{appr} \chi_a^2\approx\chi_0^2\Big(1.13+3.76\,a^2\Big)
=\left( \chi_a^{\scriptscriptstyle B}\right)^2\Big(1+3.34\,a^2\Big)
 \ea
with the so-called `Born parameter'
$a=Z\alpha/\beta\,$.
The second term in \eqref{appr} represents the deviation from the
Born approximation. If the value of this term equal to zero, the
screening angle becomes $\chi_a=\chi_a^{\scriptscriptstyle
B}=\chi_0\sqrt{1.13}$.

The angle $\chi_0$ is defined by
\ba \chi_0= 1.13\,\frac{Z^{1/3}m}{137\,
p}=\frac{Z^{1/3}m\alpha}{0.885\, p}\,, \ea
where $p=mv$ is the incident particle momentum, and $v$ is the
particle  velocity in the laboratory frame.

The characteristic angle  is defined as
\ba\label{char} \chi_c^2=4\pi n_0t\left(\frac{Z\alpha}{\beta p}
\right)^2. \ea
Its physical meaning is that the total probability of single
scattering through an angle greater than $\chi_c$ is exactly one.

Putting $\chi_c\eta=y$ and setting $\theta/\chi_c=u$, we get
Moli\`ere's transformed equation

\ba\label{simpl} W_{\scriptscriptstyle M}(\theta)\theta d\theta = u
du\int\limits_0^{\infty}y dy J_0(u y)\exp\left\{
-\frac{y^2}{4}\left[b-\ln\left(\frac{y^2}{4}\right)\right]\right\},
\ea
for the most important values of $\eta$ of order of $1/\chi_c$. This
equation is much simpler than \eqref{mol3}.

In order to obtain a result valid for large all angles,  Moli\`{e}re
defines a new parameter $B$ by the transcendental equation
\ba B-\ln B=b. \ea
The angular distribution function can then be written as
\ba\label{exp2} W_{\scriptscriptstyle M}(\theta,B)
&=&\frac{1}{\overline{\theta^{\,2}}}\int\limits_0^{\infty}y dy J_0
(\theta
y)e^{-y^2/4}\exp\left[\frac{y^2}{4B}\ln\left(\frac{y^2}{4}\right)\right].
\ea
The Moli\`{e}re expansion method is to consider the term
$[y^2\ln(y^2/4)]/4B$ as a small parameter. This allows expansion of
the angular distribution function $W_{\scriptscriptstyle M}$ in a
power series in $1/B$:
\ba\label{power} W_{\scriptscriptstyle
M}(\theta,t)&=&\sum\limits_{n=0}^{\infty}\frac{1}{n!}\frac{1}{B^n}W_n(\theta,t)
\ea
with
\ba W_n(\theta,t)
&=&\frac{1}{\overline{\theta^{\,2}}}\int\limits_0^{\infty}y dy J_0
\left(\frac{\theta}{\bar\theta}\,
y\right)e^{-y^2/4}\left[\frac{y^2}{4}\ln\left(\frac{y^2}{4}\right)\right]^n,
\ea
$$\overline{\theta^{\,2}}=\chi_c^2B=4\pi
n_0t\left(\frac{Z\alpha}{pv} \right)^2B(t).$$\\
This method is valid for $B\geq 4.5$ and
$\overline{\theta^{\,2}}<1$. The first function $W_0(\theta,t)$ has
a simple analytical form:
\ba\label{W_0}
W_0(\theta,t)=\frac{2}{\overline{\theta^{\,2}}}\exp\left(\!\!-
\frac{\theta^2}{\overline{\theta^{\,2}}}\right), \ea
\ba \overline{\theta^{\,2}} \mathop{\sim}\limits_{\;\;t\,\to
\,\infty}\;t\ln t. \ea
For small angles, i.e., $\theta/\bar \theta=
\theta/(\chi_c\sqrt{B})=\Theta$ less than about 2, the Gaussian
\eqref{W_0} is the dominant term.  In this region, $W_1(\theta,t)$
is in general less than $W_0(\theta,t)$, so that the corrections to
the Gaussian is of order of $1/B$, i.e., of order of $10\%$. An
alternative way of obtaining the approximate solution \eqref{W_0} of
\eqref{mol3} for a thick target is given in Appendix.

A good approximate representation of the distribution for any angle
is $W_0(\theta,t)+ B^{-1}W_1(\theta,t)$, where
\ba\label{W_1} W_1(\theta,t)=\frac{2}{\overline{\theta^{\,2}}}
\exp\!\left(\!\!-\frac{\theta^2}{\overline{\theta^{\,2}}}\right)\!\left\{
\left(\!\frac{\theta^2}{\overline{\theta^{\,2}}}-1\!\right)
\!\!\left[\overline{Ei}\!\left(
\frac{\theta^2}{\overline{\theta^{\,2}}} \right)\!-\ln\!\left(
\frac{\theta^2}{\overline{\theta^{\,2}}} \right) \right]\!+1 \!
\right\}-2, \ea
\ba\overline{Ei}(\Theta)=Ei(\Theta)+\pi i\ea
with the exponential integral \cite{Stegun}
\ba Ei(\Theta)=-\int\limits_{-\Theta}^{\infty}e^{-t}\frac{dt}{t}.\ea

\section{Moli\`{e}re's determination of the screening parameters }

On the one hand, Moli\`ere writes the elastic Born cross section for
the fast charged particle scattering in the atomic field as follows:
\ba\label{q} \sigma^{\scriptscriptstyle
B}(\chi)=\sigma^{\scriptscriptstyle
R}(\chi)\left(1-\frac{F_{\scriptscriptstyle
A}(p\chi)}{Z}\right)^2=\sigma^{\scriptscriptstyle
R}(\chi)\,\,q^{\scriptscriptstyle B}(\chi).\ea
\noindent For angles $\chi$ small compared with a radian, the exact
Rutherford formula has a simple approximation:
\ba\label{sigm}\sigma^{\scriptscriptstyle B}(\chi)&=&
\frac{\theta_c^2}{4\pi n_0 t
(1-\cos\chi)^2\,\chi^4}\,\,q^{\scriptscriptstyle
B}(\chi)\\\label{sig} &\approx& \frac{\theta_c^2}{\pi n_0
t\,\chi^4}\,\,q^{\scriptscriptstyle B}(\chi). \ea
Here, $F_{\scriptscriptstyle A}$ is the atomic form factor and
$q^{\scriptscriptstyle B}(\chi)$ is the ratio of actual to the
Rutherford scattering cross sections in the Born approximation.

Then the screening angle $\chi_a^{\scriptscriptstyle B}$ in the Born
approximation one can represent via $F_{\scriptscriptstyle A}$ or
$q^{\,\scriptscriptstyle B}(\chi)$ by the equations
\ba\label{def} -\ln\big(\chi^{\scriptscriptstyle B}_a\big)
&=&\lim\limits_{\varsigma\rightarrow
\infty}\left[\int\limits_0^\varsigma\left(1-\frac{F_{\scriptscriptstyle
A}(p\chi)}{Z}\right)^2\frac{
d\chi}{\chi}+\frac{1}{2}-\ln \varsigma  \right]\\
\label{defin}&=&\lim\limits_{\varsigma\rightarrow
\infty}\left[\int\limits_0^\varsigma\frac{q^{\,\scriptscriptstyle
B}(\chi)d \chi}{\chi}+\frac{1}{2}-\ln \varsigma\right]
 \ea
with an angle $\varsigma$ such as
\ba\label{zeta}\chi_0\ll \varsigma\ll 1/\eta\sim \chi_c,\ea
where $\chi_0\sim m_e\alpha Z^{1/3}/p$.

Moli\`ere's approximation for the Thomas--Fermi form factor
$F_{\scriptscriptstyle T-F}(q)$ with momentum transfer $\vec q$ can
be written as
\ba\label{F} F_{\scriptscriptstyle T-F}(q)^{\scriptscriptstyle
M}=\sum\limits_{i=1}^{3}\frac{c_i\lambda_i^2}{q^2+\lambda_i^2}\,,
\ea
\noindent in which
$$c_1=0.35,\quad c_2=0.55,\quad c_3=0.10,$$
$$\lambda_1=0.30\lambda,\quad \lambda_2=4\lambda_1,\quad \lambda_3=5\lambda_2.$$

When the Born parameter becomes zero, the equation \eqref{def} for
the screening angle can be evaluated directly, using the facts that
$q(0)=0$ and $\lim\limits_{\varsigma\rightarrow
\infty}q(\varsigma)=1$. Then with use of \eqref{q} and \eqref{F},
can also be obtained the following approximation for
$\big(\chi_a^{\,\prime}\big)^{\scriptscriptstyle B}$
\cite{M47,Scott49}:
\ba \label{theta_a}\big(\chi_a^{\,\prime}\big)^{\scriptscriptstyle
B}=\left[\exp(C_{\scriptscriptstyle
E}-0.5)\right]\,\frac{\lambda}{p}\,\,A=\sqrt{1.174}\,\,\,\chi_0\,A,\ea
where  $\lambda=m_e \alpha Z^{1/3}/0.885$. Note that a misprint is
admitted in \cite{M47,Scott49}, i.e. the factor $A=1.0825$ in
\eqref{theta_a} should be replaced by $A=1.065=\sqrt{1.13}$.

On the other hand, Moli\`ere writes the nonrelativistic Born cross
section in the form
\ba \label{sigma} \sigma^{\scriptscriptstyle B}(\chi)=
k^2\left\vert\int\limits_{0}^{\infty}\rho\, d\rho
J_0\left(2k\rho\,\sin\frac{\chi}{2}\right)\Phi_{\scriptscriptstyle
M}^{\scriptscriptstyle B}(\vec \rho)\right\vert^2 \ea
where the Born phase shift is given in units of $\hbar=c=1$ by
\ba\label{Phi_B} \Phi_{\scriptscriptstyle M}^{\scriptscriptstyle
B}(\vec \rho)=
-\frac{2}{v}\int\limits_{\rho}^{\infty}\frac{U_{\lambda}(r)dr}{\sqrt{r^2-\rho^2}}=
-\frac{1}{v}\int\limits_{-\infty}^{\infty}U_{\lambda}\left(r=\sqrt{\rho^2+z^2}
\right)dz. \ea
Here, $k$ is the wave number of the incident particle, the variable
$\rho$ corresponds to the impact parameter of the collision,
 and $U_{\lambda}(r)$ is the screened Coulomb potential of the target
atom
 \ba\label{pot} U_{\lambda}(r)=\pm
Z\,\frac{\alpha}{r}\,\Lambda(\lambda r) \ea
with Moli\`{e}re's fit to the Thomas--Fermi screening function
$\Lambda(\lambda r)$
\ba\label{fit} \Lambda(\lambda r)&\simeq& 0.1e^{-6 \,\lambda r}+
0.55e^{-1.2 \,\lambda r} +0.35e^{-0.3 \,\lambda r}.\ea

In order to obtain a result valid for large $a$ and also for large
angles $\chi$, Moli\`ere uses the WKB technique in his calculations
of the screening angle.

Exact formulas for the WKB differential cross section $\sigma(\chi)$
and the corresponding $q(\chi)$ are given in Moli\`ere's paper
\cite{M47} as follows:
\ba\label{WKB1} \sigma(\chi)=
k^2\left\vert\int\limits_{0}^{\infty}\rho \,d\rho \,J_0(k\chi
\rho)\bigg\{1-\exp\big[ i\Phi_{\scriptscriptstyle M}(\vec
\rho)\big]\bigg\}\right\vert^2,\ea
\ba\label{WKB1.5} q(\chi)=
\frac{(k\chi)^4}{4\,a^2}\left\vert\int\limits_{0}^{\infty}\rho\,
d\rho J_0(k\chi\rho)\bigg\{1-\exp \big[i\Phi_{\scriptscriptstyle
M}(\vec \rho)\big]\bigg\}\right\vert^2\ea
with the phase shift given by
\ba\label{WKB2}\Phi_{\scriptscriptstyle M}(\vec
\rho\,)=\int\limits_{-\infty}^{\infty}\Big[k_r(r)-k\Big]dz, \ea
where $k_r(r)$ is the relativistic wave number for the particle at a
distance $r$ from the nucleus, and the quantity $\rho$ is seen to be
impact parameter of the trajectory or `ray'. As before, $k$ is the
initial or asymptotic value of the wave number.

When $k_r(r)$ is expanded as a series of powers of
$U_{\lambda}(r)/k$, the first-degree term yields the same expression
for $\Phi_{\scriptscriptstyle M}(\vec \rho\,)$ as \eqref{Phi_B}. The
Born approximation for \eqref{WKB1} is obtained by expanding the
exponential in \eqref{WKB1} to first order in the Born parameter
$a$.

The relations \eqref{sig} and \eqref{defin} between the quantities
$\sigma^{\scriptscriptstyle B}(\chi)$, $q^{\scriptscriptstyle
B}(\chi)$, and  $\chi_a^{\scriptscriptstyle B}$ remain valid  for
the quantities $\sigma(\chi)$, $q(\chi)$, and $\chi_a$.

Despite the fact that the formulas \eqref{WKB1} and \eqref{WKB1.5}
are exact, evaluation of these quantities was carried out by
Moli\`{e}re only approximately. To estimate \eqref{WKB1.5},
Moli\`{e}re used the first-order Born shift \eqref{Phi_B} with
\eqref{pot} and \eqref{fit}, what is good only to terms of first
order in $a$, and he found
\ba\label{WKB3} q(\chi)\approx \bigg\vert\,
1\!&\!-\!&\!\frac{4ia(1-ia)^2}{(\chi/\chi_0)^2}
\bigg\{-0.81+2.21\,\bigg[-\Re\left[\psi(ia)\right]-\frac{1}{1-ia}+\frac{1}{2ia}+
\lg\frac{\chi}{2\chi_0}\bigg]\bigg\}\bigg\vert^2.\ea
Here, $\psi$ is the so-called `digamma function', i.e., the
logarithmic derivative of the $\Gamma$-function
$\psi(x)=d\ln\Gamma(x)/dx$.

He has fitted  a simple formula to the function
$\Re\left[\psi(ia)\right]$ from \eqref{WKB3}:
\ba\label{psi} \Re\left[\psi(ia)\right]\approx\frac{1}{4}\lg
\left(a^4+\frac{a^2}{3}+0.13\right). \ea
Inserting \eqref{psi} into \eqref{WKB3} and neglecting terms of
orders higher than $a^2$, he got
\ba\label{WKB4} q(\chi)\approx 1-\frac{8.85}{(\chi/\chi_0)^2}
\bigg[1+ 2.303\,a^2\lg\frac{7.2\cdot 10^{-4}(\chi/\chi_0)^4}
{\left(a^4+a^2/3+0.13\right)}\bigg].\ea
Moli\`{e}re has calculated $q(\chi)$ for different $a$ values. As a
result, he has devised an interpolation scheme based on a linear
relation between $(\chi/\chi_0)^{2}$ and $a^2$ for fixed $q$:
\ba (\chi/\chi_0)^{2}\approx A_q+a^2B_q. \ea
Calculating the screening angle defined by
\ba\label{defex}
-\ln\big(\chi_a\big)=\frac{1}{2}+\lim\limits_{\varsigma\rightarrow
\infty}\left[\int\limits_0^\varsigma\frac{q(\chi)d \chi}{\chi}-\ln
\varsigma\right]
=\frac{1}{2}-\ln\chi_0-\int\limits_0^1dq\ln\left(\frac{\chi}{\chi_0}\right)
\ea
and assuming a linear relation between $\chi_a^2$ and $a^2$,
Moli\`{e}re writes finally the following interpolating formula for
the screening angle:
\ba\label{interpol} \chi_a\approx\chi_0\sqrt{1.13+3.76\,a^2}. \ea
Critical remarks to his derivation of this result are given in
\cite{Scott49,nigam}.

\section{Alternative determining the screening parameters}

To obtain an exact correction to the first-order Born screening
angle $\big(\chi_a^{\,\prime}\big)^{\scriptscriptstyle B}$, we will
carry out our analytical calculation in terms of the function $
Q_(\eta) $:
\ba\label{Qexact} Q(\eta)=2\pi \int\limits_0^{\infty}
\sigma(\chi)[1-J_0(\chi \eta)]\chi d\chi \equiv\int
d^2\rho\Big[1-cos\big[\Delta \Phi(\vec \rho,\vec \eta\,)\big],
\Big]\ea
where the phase shift can be determined by the equation
\ba\label{Phiexact}
\Delta\Phi(\vec\rho,\vec\eta\,)=\Phi(\rho_+)-\Phi(\rho_-),\quad
\vec\rho_\pm=\vec\rho\pm\vec\eta/2p. \ea

Substituting the expression for the cross section
\ba\label{qq} \sigma(\chi) = \frac{\chi_c^2}{\pi n_0
t\,\chi^4}\,\,q(\chi) \ea
into \eqref{Qexact}, we rewrite it in the form:
\ba\label{Q_1,2} n_0t\,Q(\eta)=2\chi_c^2 \int\limits_0^{\infty}
[1-J_0(\chi \eta)]\, q(\chi)\chi^{-3} d\chi. \ea
For the important values of $\eta$ of order of $1/\chi_c$ or less,
it is possible to split the last integral into two integrals at the
angle $\varsigma$ \eqref{zeta}:
$$I(\eta)=\int\limits_0^{\infty} [1-J_0(\chi \eta)]\,
q(\chi)\chi^{-3} d\chi$$ $$ =\int\limits_0^{\varsigma} [1-J_0(\chi
\eta)]\, q(\chi)\chi^{-3} d\chi+ \int\limits_{\varsigma}^{\infty}
[1-J_0(\chi \eta)]\, q(\chi)\chi^{-3} d\chi$$
\ba =I_1(\varsigma\eta)+I_2(\varsigma\eta)\,. \ea
For the part from $0$ to $\varsigma$, we can write $1-J_0(\chi
\eta)=\chi^2 \eta^2/4$, and the integral $I_1$ reduces to a
universal one, independently on $\eta$:

\ba \label{int1}I_1(\varsigma\eta)=\frac{\eta^2}{4}
\int\limits_0^{\varsigma}q(\chi)\,d\chi/\chi. \ea
For the part from $\varsigma$ to infinity, the quantity $q(\chi)$
can be replaced by unity, and  the integral $I_2$ can be integrated
by parts. This leads to the following result for $I_2$:
\ba
I_2(\varsigma\eta\,)=\frac{\eta^2}{4}\bigg[1-\ln(\varsigma\eta)+\ln
2-C_{\scriptscriptstyle E}+O(\varsigma\eta)\bigg]. \ea
Integrating \eqref{int1} with the use of \eqref{defex}, substituting
obtained solutions back into \eqref{Q_1,2}, and using the definition
$$\ln \left(\chi_c/\chi_a\right)^2+1-2C_{\scriptscriptstyle E}=\ln
\left(\chi_c/\chi_a^{\,\prime}\right)^2,$$
we arrive at a result for $Q^{}(\eta)$:
\ba Q(\eta)&=&-
\frac{(\chi_c\eta)^2}{2n_0t}\left[\ln\left(\frac{\chi_c^2\eta^2}{4}\right)-\ln
\left(\frac{\chi_c}{\chi_a^{\,\prime}}\right)^2 \right]
=-\frac{(\chi_c\eta)^2}{2n_0t}
\ln\left(\frac{\eta^{\,2}\left(\chi_a^{\,\prime}\right)^2}{4}\right).
\ea
Finally, considering the definition of $\theta_c$ \eqref{char}, we
can represent $Q(\eta)$ by the following expression:
\ba \label{res} Q(\eta)=-2\pi\left(\frac{Z\alpha}{\beta
\,p}\right)^2\eta^{\,2}
\ln\left(\frac{\eta^{\,2}\left(\chi_a^{\,\prime}\right)^2}{4}\right).
\ea
Then the screening angle $\chi_a^{\,\prime}$  can be determined via
$Q(\eta)$ by a linear equation:
\ba \label{df}
-\ln\big(\chi_a^{\,\prime}\big)^2=\ln\left(\frac{\eta^{\,2}}{4}\right)
+\left[2\pi\eta^{\,2}\left(\frac{Z\alpha}{\beta
\,p}\right)^{2}\right]^{-1}Q(\eta). \ea
Let us present the quantity $Q_{el}(\eta)$ in the form:
\ba Q(\eta\,) =Q^{\scriptscriptstyle B}(\eta\,)
-\Delta_{\scriptscriptstyle CC}[Q(\eta)]. \ea
Making  use of \eqref{res}, the difference  $\Delta
_{\scriptscriptstyle CC}[Q_{el}(\eta)]<0$ between the  Born
approximate $Q_{el}^{\scriptscriptstyle B}(\eta\,)$ and exact in the
Born parameter results for the quantity $Q_{el}^{}(\eta\,)$ can be
reduced to a difference between the quantities
$\ln\big(\chi_a^{\,\prime}\big)$ and
$\ln\big(\chi_a^{\,\prime}\big)^{\scriptscriptstyle B}$:
\ba \Delta_{\scriptscriptstyle CC}[Q(\eta)]\equiv
Q^{\scriptscriptstyle B}(\eta\,)- Q(\eta\,)\nonumber\ea $$=4\pi
\eta^{\,2} \Bigg(\frac{Z\alpha}{\beta
p}\Bigg)^2\left[\ln\big(\chi_a^{\,\prime}\big)-
\ln\big(\chi_a^{\,\prime}\big)^{\scriptscriptstyle B}\right] \equiv
4\pi \eta^{\,2} \Bigg(\frac{Z\alpha}{\beta
p}\Bigg)^2\Delta_{\scriptscriptstyle
CC}[\ln\big(\chi_a^{\,\prime}\big)].$$
On the other hand, this difference can be reduced to a difference
$\Delta q(\chi)=q^{\scriptscriptstyle B}(\chi)-q(\chi)$:
\ba\label{DeltaQ} \Delta_{\scriptscriptstyle CC} [Q(\eta)]=2\pi\!\!
\int\limits_0^{\infty}\!\!\chi d\chi \Delta\sigma(\chi)[1-J_0(\chi
\eta)]
=\frac{2\chi_c^2}{n_0t}\int\limits_0^{\infty}\frac{d\chi}{\chi^{3}}
\Delta q(\chi) [1-J_0(\chi \eta)]. \ea
Using \eqref{WKB1.5} and \eqref{Phiexact}, we get for the last
integral
\ba\label{result1} \mathop{\Delta _{\scriptscriptstyle CC}
[Q(\eta\,)]}\limits_{\eta\,\rightarrow \,0}&=& 4\pi \eta^{\,2}
\Bigg(\frac{Z\alpha}{\beta \,p}\Bigg)^2
\left[\frac{1}{2}\psi\bigg(i\,\frac{Z\alpha}{\beta}\bigg)+
\frac{1}{2}\psi\bigg(\!\!\!-i\,\frac{Z\alpha}{\beta}\bigg)
-\psi(1)\right]\\
\label{result2}&=& 4\pi \eta^{\,2} \Bigg(\frac{Z\alpha}{\beta
\,p}\Bigg)^2
\left\{\Re\left[\psi\bigg(1+i\,\frac{Z\alpha}{\beta}\bigg)\right]+C_{\scriptscriptstyle
E}\right\}, \ea
where
$$\Re\left[\psi\left(1+ia\right)\right]=\Re\left[\psi\left(1-ia\right)\right]
=\Re\left[\psi\left(ia\right)\right]=\Re\left[\psi\left(-ia\right)\right]$$
\ba\label{multi}  =-C_{\scriptscriptstyle E} +
a^2\sum\limits_{n=1}^{\infty}\frac{1}{n(n^2+a^2)}=-C_{\scriptscriptstyle
E}+  f(a),\ea
$$-\infty<a<\infty,$$
$\psi(1)=-C_{\scriptscriptstyle E}$, and
$f(a)=a^2\sum\nolimits_{n=1}^{\infty}\left[n(n^2+a^2)\right]^{-1}$
is `an universal function of $a=Z\alpha/\beta$\,'.

Finally, we get the following rigorous relations between the
quantities $\ln\big(\chi_a^{\,\prime}\big)$ and
$\ln\big(\chi_a^{\,\prime}\big)^{\scriptscriptstyle
B}$\footnote{This result can also be obtained in other ways, with
use of the technique developed in \cite{TVG}.}:
\ba\label{basres2} \ln\big(\chi_a^{\,\prime}\big)-\ln
\big(\chi_a^{\,\prime}\big)^{\scriptscriptstyle B}
&=&\,\Re\big[\psi(1+ia)-\psi(1)\big ]\,, \\
\label{basres1} \Delta_{\scriptscriptstyle
CC}[\ln\big(\chi_a^{\,\prime}\big)]&=&
\,a^2\sum\nolimits_{n=1}^{\infty}[n(n^2+a^2)]^{-1}\,\,. \ea

We point out that the relations \eqref{result2}, \eqref{basres2},
and \eqref{basres1} are independent on the form of electron
distribution in atom and are valid for any atomic model.

 From
\eqref{result2} also follows an expression for the correction to the
exponent of \eqref{mol3}. Since $\ln [g(\eta)]=-n_0t\,Q$, we have:
\ba\label{comform} \Delta_{\scriptscriptstyle CC}[ \ln
g(\eta)]&\equiv& \ln [g(\eta)]- \ln [g^{\scriptscriptstyle
B}(\eta)]\noindent\ea
$$=4\pi \eta^{\,2}n_0t
\Bigg(\frac{Z\alpha}{\beta \,p}\Bigg)^2f(a). $$
For the specified value of $\eta^2=1/\chi^2_c$, we can evaluate this
correction using the definition of $\chi_c$ \eqref{char}:
\ba\label{result3} \Delta_{\scriptscriptstyle CC}[ \ln g(\chi_c)]=
\frac{4\pi n_0t}{\chi^2_c}\frac{\chi_c^2}{4\pi \,n_0t} f(a)=f(a).
\ea

The formulas for the so-called  `Coulomb corrections' (CC), defined
as a difference between the exact and the Born approximate  results,
are known as the Bethe--Bloch formulas for the ionization losses
\cite{Bloch} and the formulas for the Bethe--Heitler cross section
of bremsstrahlung \cite{Heitler}\footnote{ The more complicate
formal expression for CC was derived by I. {\O}verb{\o} in
\cite{Overbo}.}.

The similar expression was found for the total cross section of the
Coulomb interaction of compact hadronic atoms with ordinary target
atoms \cite{TVG}. Also, Coulomb corrections were obtained to the
cross sections of the elastic and quasielastic electron scattering,
the coherent electroproduction of vector mesons \cite{Aste}, the
pair production in nuclear collisions \cite{Ivanov}, as well as to
the solutions of the Dirac and Klein--Gordon equations \cite{DirEq}.

Specificity of the expressions obtained in the present work is that
they define the Coulomb corrections to the screening angle
$\big(\chi_a^{\prime}\big)^{\scriptscriptstyle B}$, the exponential
part $g(\eta,t)$ of the distribution function $W(\theta)$, and the
angular distribution. A characteristic feature of these corrections
is their positive value, in contrast to a negative value of the
Coulomb corrections to the cross sections and the energy spectrum in
the high energy region.

\section{Relative Coulomb corrections to the Born approximation}

Let us write \eqref{basres1} as follows:
\ba\label{Ourres}
\big(\chi_a^{\,\prime}\big)&=&\big(\chi_a^{\,\prime}\big)^{\scriptscriptstyle
B}\,\exp \Big[f\left(a\right)\Big].\ea
Then relative Coulomb correction to the Born screening angle
$\big(\chi_a^{\,\prime}\big)^{\scriptscriptstyle B}$ can be
represented as
\ba\label{del0}\delta_{\scriptscriptstyle
CC}\big(\chi_a^{\,\prime}\big)=\frac{\chi_a^{\,\prime}-
\big(\chi_a^{\,\prime}\big)^{\scriptscriptstyle B}}
{\big(\chi_a^{\,\prime}\big)^{\scriptscriptstyle
B}}=\frac{\Delta\big(\chi_a^{\,\prime}\big)}
{~~~~\big(\chi_a^{\,\prime}\big)^{\scriptscriptstyle
B}}=\delta_{\scriptscriptstyle CC}\big(\chi_a^{}\big) =\exp
\left[f\left(a\right)\right]-1. \ea
As follows from \eqref{result3}, the relative CC to the exponent
$g^{\scriptscriptstyle B}(\eta)$ at $\eta^2=1/\chi^2_c$ can also be
determined by this quantity: $\delta_{\scriptscriptstyle
CC}\big(\chi_a^{}\big)=\delta_{\scriptscriptstyle
CC}\left[g(\chi_c)\right]$. Moreover, because
\ba\label{mend} \Delta_{\scriptscriptstyle
CC}\left[W(\chi_c,t)\right]\equiv W_{\scriptscriptstyle M}-
W^{\scriptscriptstyle B}_{\scriptscriptstyle M}=
\int\limits_0^{\infty} J_0(\theta \eta)\Delta g(\chi_c) \eta d\eta,
\ea
accounting  for $\int\nolimits_0^{\infty}d\eta\,\eta\,
J_{0}(\theta\eta)=0$, we get
\ba\label{mendW} \delta_{\scriptscriptstyle
CC}\left[W_{\scriptscriptstyle
M}(\chi_c,t)\right]=\frac{\Delta_{\scriptscriptstyle CC}
\left[W(\chi_c,t)\right]} {W^{\scriptscriptstyle
B}(\chi_c,t)_{\scriptscriptstyle
M}}=\frac{\Delta_{\scriptscriptstyle CC}\left[g(\chi_c)\right]}
{g^{\scriptscriptstyle B}(\chi_c)} =\exp
\left[f\left(a\right)\right]-1. \ea
Thus,
\ba\label{uni}\delta_{\scriptscriptstyle
CC}\equiv\delta_{\scriptscriptstyle CC}\big(\chi_a^{}\big)=
\delta_{\scriptscriptstyle CC}\left[g(\chi_c)\right]=
\delta_{\scriptscriptstyle CC}\left[W_{\scriptscriptstyle
M}(\chi_c,t)\right]=\exp \left[f\left(a\right)\right]-1.\ea
The numerical values of this correction are presented in Table 1.
Figure 1 illustrates their $Z$ dependence.

Let us notice that the following equivalent to \eqref{WKB3} equation
\ba\label{molcorr} q(\chi)\approx 1-\frac{8.85}{(\chi/\chi_0)^2}
\bigg\{1+4a^2\, \left[\lg\left(\frac{\chi}{2\chi_0}\right)
-f\left(a\right)-0.543\right] \bigg\}\ea
yields an approximate expression for the relative correction
$\delta(\sigma)=(\sigma-\sigma^{\scriptscriptstyle
R})/\sigma^{\scriptscriptstyle R}$ to the Rutherford cross section:
\ba\label{corrtoruth} \delta(\sigma)\approx
\frac{8.85}{(\chi/\chi_0)^2} \bigg\{1+4a^2\,
\left[\lg\left(\frac{\chi}{2\chi_0}\right)
-f\left(a\right)-0.543\right] \bigg\}.\ea
The inner part of this expression  is close in the form to the
insides of the formulas' of Bethe--Bloch \cite{Bloch},
Bethe--Maximon \cite{Heitler}, and the formula's for the total cross
section obtained in \cite{TVG}.

In order to estimate the accuracy of the Moli\`{e}re theory in
determining the Coulomb correction to the screening angle
$\chi_a^{}$, we define the difference
 and relative difference between the values of
$\delta^{}_{\scriptscriptstyle M}\big(\chi_a^{}\big)$ and
$\delta_{\scriptscriptstyle CC}\big(\chi_a^{}\big)$ by the relation
\ba\label{corr3}\delta_{\scriptscriptstyle
CCM}(\delta_{\scriptscriptstyle
CC})=\frac{\Delta_{\scriptscriptstyle
CCM}(\delta_{\scriptscriptstyle CC})}{\delta^{}_{\scriptscriptstyle
M}\big(\chi_a^{}\big)} =-\frac{\delta^{}_{\scriptscriptstyle
CC}\big(\chi_a^{}\big)-\delta_{\scriptscriptstyle
M}\big(\chi_a^{}\big)}{\delta^{}_{\scriptscriptstyle
M}\big(\chi_a^{}\big)}=1-\frac{\delta_{\scriptscriptstyle
CC}\big(\chi_a^{}\big)}{\delta^{}_{\scriptscriptstyle
M}\big(\chi_a^{}\big)}, \ea
where
\ba\label{mol} \delta^{}_{\scriptscriptstyle
M}\big(\chi_a^{}\big)=\frac{\chi^{}_a-\chi_a^{\scriptscriptstyle
B}}{\chi_a^{\scriptscriptstyle B}}=\sqrt{1+3.34}-1. \ea

To estimate the accuracy of the Moli\`{e}re theory in determining
the screening angle itself by the following relative difference
between the  approximate $\chi_a^{\scriptscriptstyle M}$ and exact
$\chi_a$ results
\begin{eqnarray}\label{angle}
\delta_{\scriptscriptstyle CCM}(\chi_a)&\equiv&\frac{\chi_a-
\chi_a^{\scriptscriptstyle M}}{\chi_a^{\scriptscriptstyle
M}}=\frac{\chi_a}{\chi_a^{\scriptscriptstyle M}}-1,
\end{eqnarray}
we rewrite (\ref{del0}) and (\ref{mol}) as
$\delta_{\scriptscriptstyle
CC}(\chi_a)+1=\chi_a/\chi_a^{\scriptscriptstyle B}$ and
$\delta^{}_{\scriptscriptstyle M}(\chi_a)+1=\chi^{\scriptscriptstyle
M}_a/\chi_a^{\scriptscriptstyle B}$. As a result, we obtain the
expression
\begin{eqnarray}
\label{rat} \delta_{\scriptscriptstyle
CCM}(\chi_a)=\frac{\Delta_{\scriptscriptstyle
CCM}(\delta_{\scriptscriptstyle CC})}{\delta^{}_{\scriptscriptstyle
M}(\chi_a)+1}.
\end{eqnarray}

In order to obtain the numerical results for the above Coulomb
corrections $\Delta_{\scriptscriptstyle
CC}\left[\ln\big(\chi_a^{\,\prime}\big)\right]=\Delta_{\scriptscriptstyle
CC}\left[\ln g(\chi_c)\right]=f(a)>0$,  $\delta_{\scriptscriptstyle
CC}\equiv\delta_{\scriptscriptstyle CC}\big(\chi_a^{}\big)=
\delta_{\scriptscriptstyle CC}\left[g(\chi_c)\right]=
\delta_{\scriptscriptstyle CC}\left[W_{\scriptscriptstyle
M}(\chi_c,t)\right]>0$, and $\delta_{\scriptscriptstyle
CCM}[\chi_a]$, according to \eqref{basres1}, \eqref{del0}, and
\eqref{rat},  we must first calculate the values of the function
$f(a)=\Re\big[\psi(1+ia)\big ]+C_{\scriptscriptstyle E}$.

\newpage

\begin{center}

{\bf Table 1.} The $Z$ dependence of the  corrections and the
differences defined by \eqref{del0}, \eqref{corr3}, \eqref{mol},
\eqref{rat}, \eqref{f(a)}, and \eqref{sum}.

\end{center}

\begin{center}

\begin{tabular}{lcccccccc}
\hline \\[-2mm] $M~~$&$~~Z~~$&$~~\delta^{}_{\scriptscriptstyle CC}(\chi_a)~~$&
$~~\sum\nolimits_{n=1}^{\infty}~~$&$~~f(Z\alpha)~~$&
$~~\delta_{\scriptscriptstyle M}(\chi_a)
~$&$\Delta_{\scriptscriptstyle CCM}(\delta_{\scriptscriptstyle
CC})$& $\delta_{\scriptscriptstyle CCM}(\delta_{\scriptscriptstyle
CC}) $&$\delta_{\scriptscriptstyle CCM}(\chi_a)
$\\[.2cm]
\hline\\[-2mm]
Be&~4 &0.0010&1.2012&0.0010&0.0014&0.0004&0.2989&0.0004\\
Al&13&0.0108&1.1928&0.0107&0.0149&0.0041&0.2764&0.0040\\
Ti&22&0.0308&1.1758&0.0303&0.0422&0.0114&0.2701&0.0109\\
Ni&28&0.0499&1.1602&0.0487&0.0678&0.0179&0.2646&0.0168\\
Mo&42&0.1103&1.1127&0.1046&0.1463&0.0360&0.2459&0.0314\\
Sn&50&0.1544&1.0799&0.1436&0.2018&0.0473&0.2345&0.0396\\
Ta&73&0.3175&0.9710&0.2758&0.3959&0.0784&0.1981&0.0562\\
Pt&78&0.3590&0.9467&0.3067&0.4430&0.0840&0.1895&0.0582\\
Au&79&0.3670&0.9414&0.3125&0.4520&0.0850&0.1880&0.0585\\
Pb&82&0.3930&0.9262&0.3316&0.4820&0.0890&0.1846&0.0600\\
U& 92&0.4845&0.8761&0.3951&0.5830&0.0985&0.1689&0.0622\\[.2cm]

\hline
\end{tabular}

\end{center}

\vspace{.5cm}

From the digamma series \cite{Stegun}
\ba \psi(1+a)=1-C_{\scriptscriptstyle
E}-\frac{1}{1+a}+\sum\limits_{n=2}^{\infty}(-1)^{n}\big[\zeta(n-1)\big]a^{n-1},
\quad \vert a\vert<1, \ea
where $\zeta$ is the Riemann zeta function, leads the corresponding
power series for $\Re\big[\psi(1+ia)\big ]=\Re\big[\psi(ia)\big ]$

\ba \Re\big[\psi(ia)\big ]=1-C_{\scriptscriptstyle
E}-\frac{1}{1+a^2}+\sum\limits_{n=1}^{\infty}(-1)^{n+1}\big[\zeta(2n+1)\big]a^{2n},
\quad \vert a\vert<2, \ea
and the function
\ba\label{summa}
f(a)=a^2\sum\limits_{n=1}^{\infty}\frac{1}{n(n^2+a^2)} \ea
can be represented as follows \cite{ryzh}:
\ba
~~~~~~f(a)&=&1-\frac{1}{1+a^2}+\sum\limits_{n=1}^{\infty}(-1)^{n+1}\big[\zeta(2n+1)\big]
a^{2n},\quad \vert a\vert<2,\nonumber\\
&=&1-\frac{1}{1+a^2}+0.2021\,a^2-0.0369\,a^4+0.0083\,a^6-\ldots\label{f(a)}
\ea

An equivalent way to estimate $f(a)$  to four decimal figures is to
present the sum from \eqref{summa} in the following form
\cite{Heitler}:
\ba
\sum\limits_{n=1}^{\infty}[n(n^2+a^2)]^{-1}&=&\big(1+a^2\big)^{-1}+\sum\limits_{n=1}^{\infty}
\big(-a^2\big)^{n-1}\big[\zeta(2n+1)-1\big],\nonumber\\
\label{sum}&=&\big (1+a^2\big )^{-1}+0.20206-0.0369a^2+ 0.0083a^4
-0.002a^6.\ea
Eq. (\ref{sum}) is sufficient to evaluate  this sum up to
$a<2/3=0.667$.

The calculation results for the sum \eqref{sum}, the function $f(a)$
\eqref{f(a)}, the relative Coulomb correction
$\delta^{}_{\scriptscriptstyle CC}$ \eqref{uni}, its difference with
the Moli\`{e}re correction $\delta^{}_{\scriptscriptstyle M}$, and
the relative difference in determining the screening angle
$\delta^{}_{\scriptscriptstyle CCM}(\chi_a)$ are given in Table 1.
Some results from Table 1 are presented by Figure 1.

\begin{figure}[h]

\begin{center}

\includegraphics[width=0.78\linewidth]{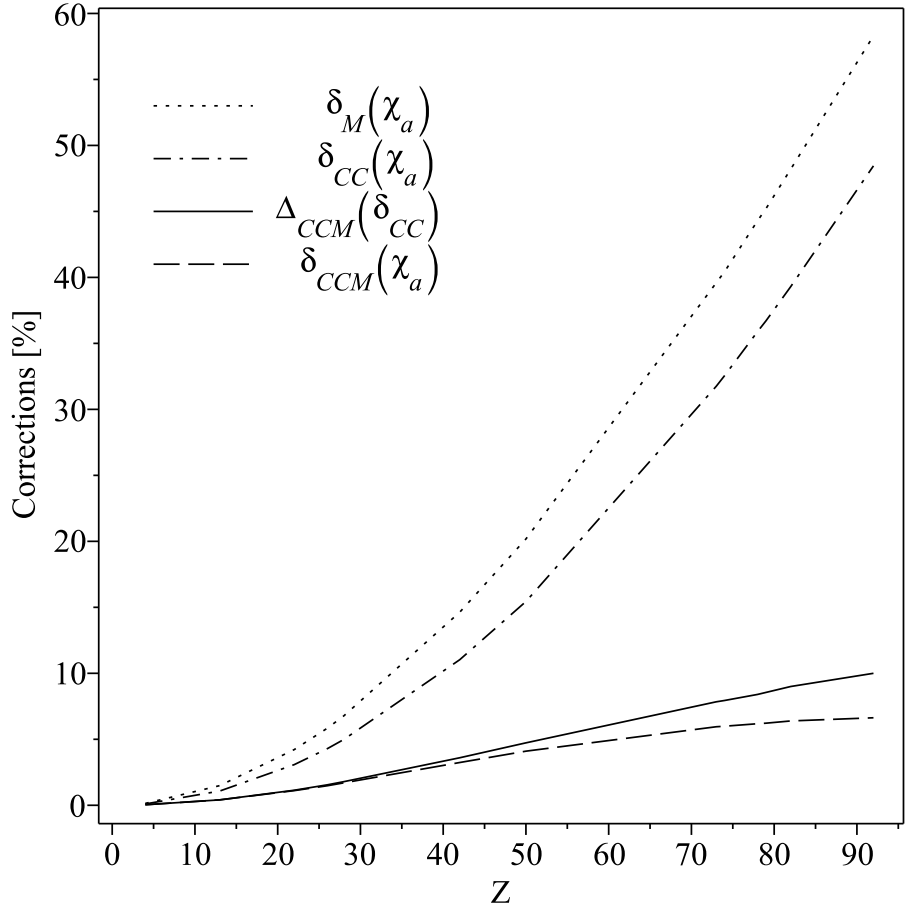}

\caption{ The dependence of the relative Moli\`{e}re  and Coulomb
corrections, as well as their difference and the relative difference
in determining the screening angle on the nuclear charge $Z$.}
\label{Fig1}

\end{center}

\end{figure}

\newpage

The Table 1 shows that the $f(Z\alpha)$ values, obtained on the
basis of (\ref{f(a)}) and (\ref{sum}), coincide up to four decimal
digits and show good agreement with the corresponding values of this
function from paper \cite{Hencken}. So $f(Z\alpha)=0.3129$
\cite{Hencken} and $f(Z\alpha)=0.3125$ (Table 1) for $Z=79$;
$f(Z\alpha)=0.3318$ \cite{Hencken} and $f(Z\alpha)=0.3316$ (Table 1)
for $Z=82$. The maximum value of the relative Coulomb correction
$\delta_{\scriptscriptstyle CC}$ amounts approximately to 50\% for
$Z=92$.

In \cite{nigam} it was found that the deviation of the screening
angle from the first Born approximation is much smaller than this
effect determined by Molier\`{e}s expression for this quantity. Our
results confirm this conclusion (Figure 1).

From Table 1 and Figure 1 it is obvious that the absolute inaccuracy
$\Delta_{\scriptscriptstyle CCM}(\delta_{\scriptscriptstyle CC})$ of
the Moli\`{e}re theory in determining the relative Coulomb
correction to the screening angle increases up to $10\%$ with the
rise of $Z$, and the corresponding relative inaccuracy
$\delta_{\scriptscriptstyle CCM}(\delta_{\scriptscriptstyle CC})$
varies between 17 and 30\% over the range $4\leq Z \leq 92$; the
$\delta_{\scriptscriptstyle CCM}(\chi_a)$ value reaches about $6\%$
for high $Z$ targets.

Thus, we can conclude that the such  large Coulomb corrections as
$\Delta_{\scriptscriptstyle CC}\equiv\Delta_{\scriptscriptstyle
CC}\left[\ln\big(\chi_a^{\,\prime}\big)\right]=\Delta_{\scriptscriptstyle
CC}\left[\ln g(\chi_c)\right]=f(a)$ and $\delta_{\scriptscriptstyle
CC}\equiv\delta_{\scriptscriptstyle CC}\big(\chi_a^{}\big)=
\delta_{\scriptscriptstyle CC}\left[g(\chi_c)\right]=
\delta_{\scriptscriptstyle CC}\left[W_{\scriptscriptstyle
M}(\chi_c,t)\right]=\exp[f(a)]-1$ should be taken into account in
describing the high-energy experiments with nuclear targets. The
accuracy of the Moli\`{e}re theory in determining the Coulomb
correction to the screening angle and the screening angle itself
must also be taken into consideration.

\newpage

\section{Summary and Conclusions}

\begin{enumerate}

\item We obtained the rigorous relations between Born and
the exact values of the quantities $Q(\eta)$, $\ln\left[
g(\eta)\right]$, and $\chi_a^{\,\prime}$, which do not depend on the
shape of the electron density distribution in the atom and are valid
for any atomic model. The main limitation of the presented exact
results consists in their applicability for small scattering angles.

\item Also, we evaluated numerically the Coulomb corrections
$\Delta_{\scriptscriptstyle CC}\equiv\Delta_{\scriptscriptstyle
CC}\left[\ln\big(\chi_a^{\,\prime}\big)\right]=\Delta_{\scriptscriptstyle
CC}\left[\ln g(\chi_c)\right]=f(a)$ and relative Coulomb corrections
$\delta_{\scriptscriptstyle CC}\equiv\delta_{\scriptscriptstyle
CC}\big(\chi_a^{}\big)= \delta_{\scriptscriptstyle
CC}\left[g(\chi_c)\right]= \delta_{\scriptscriptstyle
CC}\left[W_{\scriptscriptstyle M}(\chi_c,t)\right]=\exp[f(a)]-1$ for
nuclear charge ranged from $Z=4$ to $Z=92$.

\item We found that these Coulomb corrections have a large value for
high Z targets. For instance, the magnitude of
$\delta_{\scriptscriptstyle CC}\left(\chi_a^{}\right)$ is about
$40\div 50\%$ for $Z\sim 80\div 90$. The contribution of such
corrections is larger than experimental errors in the most high
energy experiments whose measurement accuracy has an order of a few
percent, and these corrections should be appropriately considered in
experimental data processing.

\item We estimated numerically the difference and relative differences
between our results and those of Moli\`{e}re over the range $4\leq Z
\leq 92$, and we found that while the values of
$\delta_{\scriptscriptstyle CCM}(\chi_a)$ and
$\Delta_{\scriptscriptstyle CCM}(\delta_{\scriptscriptstyle CC})$
increase with Z up to $6\%$ and $10\%$, respectively, the relative
difference $\delta_{\scriptscriptstyle
CCM}(\delta_{\scriptscriptstyle CC})$ varies between 17 and 30\%
over the range $4\leq Z \leq 92$. Thus, we can conclude that these
corrections to the approximate Moli\`{e}re result must also be taken
into account  for a rather accurate description of high energy
experiments with nuclear targets.

\end{enumerate}

\section*{Acknowledgments}

One of the authors is grateful to Dr. Leonid Afanasyev, who
initiated the consideration of the problem discussed in this paper.

\section*{Appendix: Approximate solution for the thick targets}

{\small

We can  obtain the approximate solution \eqref{W_0} of \eqref{mol3}
for a thick target  in the following simple way. When the total
number of collisions is
\ba \label{thick1}N_0= 2\pi n_0
t\int\limits_0^{\infty}\sigma(\chi)\chi d\chi\gg 1, \ea
we can write
\ba \label{thick2} 1-J_0(\chi \eta)\approx \frac{\chi^2
\eta^2}{4}\ea
for small angles like $\chi_0 \eta \ll 1$. This allows one to reduce
the integral \eqref{mol3} to a much simpler one:
\ba\label{thick} W_{\scriptscriptstyle M} (\theta,t)=
\int\limits_0^{\infty}\eta\, d\eta J_0(\theta \eta)\exp\left[-2\pi
n_0 t \frac{\eta^2}{4}\int \limits_0^{\infty}\sigma(\chi)\chi^3
d\chi\right]. \ea
Since
\ba\lim\limits_{\chi\rightarrow \infty}\sigma(\chi)\chi^3
\rightarrow 0,\ea
the corresponding integrand from \eqref{thick} is a convergent
integral
\ba\int\limits_0^{\infty}\sigma(\chi)\chi^3 d\chi < \infty.\ea
Taking into account
%
\ba\label{Gamma} \int\limits_0^{\infty}d\eta\,\eta\,
J_{0}(\theta\eta)=2c^{-2}\,\frac{\Gamma\left(1\right)}{\Gamma\left(0\right)}=0\ea
with the Gamma function $\Gamma(x)=(x-1)!$ \cite{ryzh}, we get a
final result for \eqref{thick}:
\ba\label{approx} W_{\scriptscriptstyle
M}(\theta,t)\approx\frac{2}{\overline{
\theta^{\,2}}}\exp\left(-\frac{\theta^2}{\overline{\theta^{\,2}}}\right),
\ea
where
\ba\label{bartheta} \overline{\theta^{\,2}}=2\pi n_0t\int
\sigma(\chi)\chi^3 d\chi. \ea
For the Rutherford law
\ba \sigma^{\scriptscriptstyle R}(\chi)=\left(\frac{2Z\alpha}{\beta
p}\right)^2\frac{1}{\chi^4},\ea
when $\sigma^{\scriptscriptstyle R}(\chi)\gg \theta_0=\chi_0$, the
quantity \eqref{bartheta} takes a value
\ba \overline{\theta^{\,2}}=2\pi n_0t\int
\sigma(\chi)\chi^3d\chi=\infty,\ea
and the approximate solution \eqref{approx} is not applicable.

}


{\small

\begin{thebibliography}{99}


\bibitem{Dirac} DIDAC-Collaboration: B. Adeva et al.,
{\it Phys. Lett.} {\bf B 704} (2011) 24; MuScat Collaboration: D.
Attwood et al., {\it Nucl. Instrum. Meth.} {\bf B 251} 41 (2006) 41;
C.M. Ankebrandt et al., Proposal of the MUCOOL Collaboration, April
2012; CERN-NA63 Collaboration: H.D. Thomsen et al., {\it Phys.
Lett.} {\bf B 672} (2009) 323; IceCube Collaboration: Aartsen M.G.
et al., {\it Phys. Rev. Lett.} {\bf 111} (2013) 021103.

\bibitem{GS} E.J. Williams, {\it Proc. Roy. Soc.}  {\bf 169 A} (1939) 531;
S. A. Goudsmit, J. L. Saunderson,  {\it Phys. Rev.} {\bf 57} (1940)
24, {\bf 58} (1940) 36; B. Rossi and K. Greisen, {\it Rev. Mod.
Phys.} {\bf 13} (1941) 240; H.W. Lewis, {\it Phys. Rev.} {\bf 78}
(1950) 526.

\bibitem{M47} G. Moli\`{e}re, {\it Z. Naturforsch.} {\bf 2 a} (1947)
133, {\bf 3 a} (1948) 78, {\bf 10 a} (1955) 177.

\bibitem{Bethe} H.A. Bethe, {\it Phys. Rev.} {\bf 89} (1953) 256.

\bibitem{Scott49} H. Snyder and W.T. Scott, {\it Phys. Rev.} {\bf 76} (1949)
220; W.T. Scott, {\it Phys. Rev.} {\bf 85} (1952) 245.

\bibitem{nigam} B.P. Nigam, M.K. Sundaresan, and T.Y. Wu, {\it Phys. Rev.} {\bf
115} (1959) 491.

\bibitem{Stegun} {\it Handbook of Mathematical Functions}, Eds. M. Abramowitz
and I.A. Stegun, National Bureau of Standards, Applied Mathematics
Series, 1964.

\bibitem{TVG} O.O. Voskresenskaya, S.R. Gevorkyan, and
A.V. Tarasov, {\it Phys. Atom. Nucl}. {\bf 61} (1998) 1517.


\bibitem{Bloch} H.A. Bethe, {\it Z. Phys.} {\bf 76} (1932) 293;
F. Bloch, {\it Ann. Phys.} {\bf 5} (1933) 285.

\bibitem{Heitler} H.A. Bethe and W. Heitler,
{\it Proc. Roy. Soc. (London)} {\bf 146 A} (1934) 83; H.A. Bethe and
L.C. Maximon, {\it Phys. Rev.} {\bf 93} (1954) 768, 788.

\bibitem{Overbo} I. {\O}verb{\o}, K.J.  Mork, and H.A. Olsen, {\it Phys. Rev.}
{\bf 175} (1968) 1978, {\it Phys. Rev.} {\bf A 8} (1973) 668; I.
{\O}verb{\o}, {\it Phys. Lett.}  {\bf B 71} (1977) 412.


\bibitem{Aste} J. Arrington, {\it J. Phys.} {\bf G 40} (2013) 115003;
A. Aste, {\it Nucl. Phys.} {\bf A 806} (2008) 191; A. Aste, C. von
Arx, and D. Trautmann, {\it Eur. Phys. J.} {\bf A 26} (2005) 167; A.
Aste and J. Jourdan, {\it Europhys. Lett.} {\bf 67} (2004); A. Aste,
K. Hencken, J. Jourdan et al., {\it Nucl. Phys.} {\bf A 743} (2004)
259; A. Aste, K. Hencken, and D. Trautmann, {\it Eur. Phys. J.} {\bf
A 21} (2004) 161.

\bibitem{Ivanov} D. Ivanov and K. Melnikov, {\it Phys. Rev.} {\bf D 57} (1998)
4025; D. Ivanov, A. Schiller, and V. Serbo, {\it Phys. Lett.} {\bf B
454} (1999) 155; A.J. Baltz, F. Gelis, L. McLerran et al., {\it
Nucl. Phys.} {\bf A 695} (2001) 395; R.N. Lee and A.I. Milstein,
{\it JETP} {\bf 109} (2009) 968.



\bibitem{DirEq} R.N. Lee and A.I. Milstein, {\it Phys. Rev.}
{\bf A 61} (2000) 032103, {\it Phys. Rev.} {\bf A 64} (2001) 032106,
{\it JETP} {\bf 104} (2007) 423; J.A. Tjon, and S.J. Wallace,
DOE/ER/40762-373 (2006).

\bibitem{ryzh} I.S. Gradshtein and I.M. Ryzhik, {\it Table of Integrals,
Series and Products}, Nauka Publication, Moscow, 1971.

\bibitem{Hencken} U.D. Jentschura, K. Hencken, and V.G. Serbo, {\it Eur. Phys.
J.} {\bf C 58} (2008) 281.


\end{thebibliography}

}
\end{document}